\RequirePackage{ifpdf}
\ifpdf 
\documentclass[pdftex]{sigma}
\else
\documentclass{sigma}
\fi

\begin{document}

\allowdisplaybreaks

\renewcommand{\thefootnote}{$\star$}

\renewcommand{\PaperNumber}{052}

\FirstPageHeading

\ShortArticleName{Symmetry and Separation in Surface and Curve Flows}

\ArticleName{The Role of Symmetry and Separation \\ in Surface Evolution and Curve Shortening\footnote{This paper is a
contribution to the Special Issue ``Symmetry, Separation, Super-integrability and Special Functions~(S$^4$)''. The
full collection is available at
\href{http://www.emis.de/journals/SIGMA/S4.html}{http://www.emis.de/journals/SIGMA/S4.html}}}

\Author{Philip BROADBRIDGE~$^\dag$ and Peter VASSILIOU~$^\ddag$}

\AuthorNameForHeading{P.~Broadbridge and P.~Vassiliou}

\Address{$^\dag$~School of Engineering and Mathematical Sciences,  La Trobe University,\\
\hphantom{$^\dag$}~Melbourne, Victoria, Australia}
\EmailD{\href{mailto:P.Broadbridge@latrobe.edu.au}{P.Broadbridge@latrobe.edu.au}}

\Address{$^\ddag$~Faculty of Information Sciences and Engineering,  University of Canberra,\\
\hphantom{$^\dag$}~Canberra, A.C.T., Australia}
\EmailD{\href{mailto:Peter.Vassiliou@canberra.edu.au}{Peter.Vassiliou@canberra.edu.au}}

\ArticleDates{Received January 23, 2011, in f\/inal form May 25, 2011;  Published online June 01, 2011}

\Abstract{With few exceptions, known explicit solutions of the curve shortening f\/low (CSE) of a plane curve, can be constructed by classical Lie  point symmetry reductions or by functional separation of variables.  One of the functionally separated solutions is the exact curve shortening f\/low of a closed, convex ``oval''-shaped curve and another is the smoothing of an initial periodic curve that is close to a square wave. The types of anisotropic evaporation coef\/f\/icient are found for which the evaporation-condensation evolution does or does not have solutions that are analogous to the basic solutions of the CSE, namely the grim reaper travelling wave, the homothetic shrinking closed curve and the homothetically expanding grain boundary groove. Using equivalence classes of anisotropic dif\/fusion equations, it is shown that physical models of evaporation-condensation must have a dif\/fusivity function that decreases as the inverse square of large slope. Some exact separated solutions are constructed for  physically consistent anisotropic dif\/fusion equations.}


\Classification{35A30; 35K55; 58J70; 74E10}

\renewcommand{\thefootnote}{\arabic{footnote}}
\setcounter{footnote}{0}

\section{Introduction}

The standard curve shortening f\/low is a nonlinear evolution by curvature,
\begin{gather}\label{CSF_intro}
 \frac{\partial \gamma}{\partial t}(x,t)=\kappa(x,t)\mathbf{n}(x,t),\qquad
 \gamma(x,0)=\gamma_0(x),
\end{gather}
where $\kappa(x,t)$ is the Euclidean curvature of $\gamma(x,t)$ and $\mathbf{n}(x,t)$ is the outward unit normal to the curve at each point.  For the curve-shortening f\/low parameterized  as $x\mapsto (x,y(x,t))$,
\begin{gather}
\mathbf{n} \cdot \mathbf{e}_y  y_t  =   \mathbf{n} \cdot \frac{\partial \gamma}{\partial t}(x,t),\qquad
\label{CSE_intro}
y_t = \frac{y_{xx}}{1+y_x^2},
\end{gather}
which describes the curve shortening f\/low  in Cartesian coordinates.  This equation arises in the practical context of metal surface evolution \cite{Mullins}. More recently, it has been used extensively as an isotropic curve-smoothing mechanism in image processing \cite{Malladi, Olver}; see also~\cite{Cao}. Equation~(\ref{CSF_intro}) is  well known and has been the subject of several important studies, for instance~\cite{GageHamilton,Grayson1}.  Few explicit and exact curve shortening f\/lows of plane curves are known in the literature.  Some similarity solutions with prescribed-slope boundary conditions have been constructed parametrically in terms of integrals of algebraic functions of elementary functions~\cite{Broadbridge, Ishimura}. King \cite{King} writes down a number non-invariant solutions. In addition,  there are known to be various self-similar rotating spiral and f\/lower-head solutions \cite{Abresch, Halldorsson}.

\looseness=-1
We demonstrate in this paper how for most  solutions known to us, Lie symmetries of various types or separable coordinate systems play a vital role and we conjecture that all currently known solutions and many new solutions can be obtained via a series of higher order constraints in the method of functional separation. In Section~\ref{section2} we brief\/ly review the solutions that can be obtained by symmetry reduction. In Section~\ref{section3}, we show  how the method of functional separation of variables recovers additional interesting explicit, non-self-similar solutions. Some of the properties and materials science applications of these solutions are  developed in more detail.  In Section~\ref{section4}, an anisotropic version of the second-order Mullins equation is derived for materials that include a realistic dependence of evaporation coef\/f\/icient on surface orientation. Equivalence classes for anisotropic evaporation coef\/f\/icients are constructed under the Euclidean group. Using the equivalence classes, in Section~\ref{section5} we  investigate under which types of anisotropy, analogs of the standard solutions for isotropic dif\/fusion, do or do not exist. Some examples of exact solutions are produced by direct construction and in Section~\ref{section6} by functional separation of variables.

\section{Symmetry reductions of the curve shortening equation}\label{section2}

One of the most widely used techniques for constructing explicit, exact solutions of nonlinear partial dif\/ferential equations is Lie symmetry reduction. This has been extensively investigated in relation to the partial dif\/ferential equation (\ref{CSE_intro}) or its derivative forms
\begin{gather}
u_t= \partial_x[D(u)u_x], \qquad  D(u)=1/\left(1+u^2\right),
\label{gendiff} \\
\kappa_t= \kappa^2\kappa_{\theta\theta}+\kappa^3,
\label{curvature} \\
R_t= \partial_{\theta}[R^{-2}R_{\theta}]-R^{-1},
\label{radius}
\end{gather}
where $u=y_x$, $\kappa$ is curvature, $R=1/\kappa$ and $\theta$ is the orientation angle along a convex curve~\cite{Angenent}. Although the standard nonlinear dif\/fusion equation~(\ref{gendiff}) is rarely used in the context of curve shortening, it has  non-trivial Lie potential symmetries that enable one to construct exact similarity solutions \cite{Reid}. The reaction-dif\/fusion equation in standard form~(\ref{radius}), another equation that is rarely used  in the context of curve shortening,  has been fully classif\/ied by Lie point symmetry reductions~\cite{Galak}.

In common with all autonomous nonlinear dif\/fusion equations of second order, the curve shortening equation (\ref{CSE_intro}) is invariant under translations in $x$, $y$ and $t$, plus the Boltzmann scaling group generated by $x\partial/\partial x + y\partial y + 2t\partial/\partial t$:
\begin{gather*}
( \bar x,  \bar{y},  \bar{t})=\big( e^\epsilon x,  e^\epsilon y,  e^{2\epsilon}t \big).
\end{gather*}
This allows the possibility of two types of scale-invariant similarity solution: the expanding solution of the form
$y/\sqrt{t-t_0}=G(x/\sqrt{t-t_0} )$, $(t >t_0)$ and the contracting solution of the form $y/\sqrt{t_0-t}=G(x/\sqrt{t_0-t})$,  $(t<t_0)$. Being Euclidean-invariant, (\ref{CSE_intro}) has the rotation group as an additional symmetry. Some interesting exact solutions to this equation, and consequently to~(\ref{gendiff}) can be constructed by consecutive symmetry reductions~\cite{Reid}; see also~\cite{ChouLi}.

\begin{figure}[t]
\centering
\includegraphics[width=9cm]{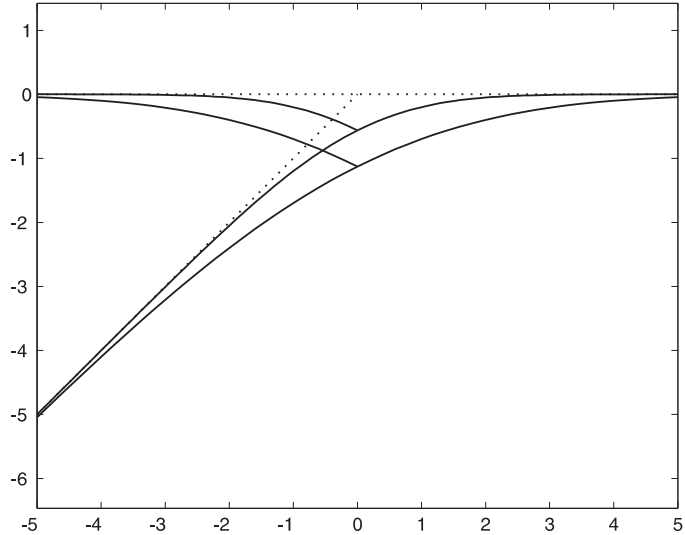}
\caption{Self-similar evolution by evaporation near a grain boundary groove (symmetrized upper curve) or equivalently by condensation in an obtuse-angled wedge. Output times are at $t=1$ and $t=4$.}
\label{fig-1}
\end{figure}

The expanding homothetic similarity solution with initial-boundary conditions $y_x(0,t)=m$, $y\to 0$, $x\to\infty$ and $y(x,0)=0$, was given in \cite{Broadbridge}. The solution takes the form of the symmetrized (upper) curve in Fig.~\ref{fig-1}, which represents an evolving grain boundary groove \cite{Mullins, Broadbridge}.

Alternatively, the above solution may be extended smoothly to the interior of an obtuse-angled wedge, as shown. These and the homothetic solutions in an acute-angled wedge, ob\-tainab\-le from the same type of reduction, make up the ``open-angle'' solutions~\cite{Ishimura}.

The  homothetically shrinking simple closed-curve solution is the circle $r=\sqrt 2 (t_0-t)^{1/2}$. Self-intersecting closed-curve ``f\/lower-head'' solutions were given in~\cite{Abresch}.

Steady-state solutions, invariant under time translations, are simply the straight lines. Likewise, the solutions that are invariant under spatial translations in a particular direction are simply the straight lines in that direction. Another simple known solution is the travelling wave solution of the form $y-ct=G(x)$, which is invariant under the translation in space and time generated by $\partial/\partial t +c \partial/\partial y$. This is the well-known ``Calabi grim reaper'' \cite{Grayson1}, taking the appearance  of a shepherd's crook
\begin{gather*}
y-ct=\frac{-1}{c}\log \cos (cx).
\end{gather*}

Uniformly rotating solutions of the form $r=G(\phi-\sigma t)$ ($r$ and $\phi$ being plane polar coordinates) follow from reduction by the symmetry generated by $\sigma (-y\partial/\partial x+x\partial/\partial y) -\partial/\partial t$.

Expanding or contracting rotating  solutions of the invariant form $r/\sqrt{t}=G(\phi-\frac{\sigma}{2a_3}\log t)$ or
$r/\sqrt{-t}=G(\phi-\frac{\sigma}{2a_3} \log(-t))$ follow from reductions under the one-parameter group generated by a linear combination of generators of dilatations and rotations
\begin{gather*}
\Gamma=\Gamma_4+\sigma \Gamma_5,\qquad
\Gamma_4=x\frac{\partial}{\partial x}+y\frac{\partial}{\partial y}+2t\frac{\partial}{\partial t},\qquad
\Gamma_5=-y\frac{\partial}{\partial x}+x\frac{\partial}{\partial y} .
\end{gather*}

A classif\/ication of the types of solutions of the associated reduced ordinary dif\/ferential equations, but without solving them explicitly and without referring to symmetries, was given recently by Halldorsson \cite{Halldorsson}.

\subsection{Reciprocal transformations}

It is well known that the class of nonlinear heat equations is stabilised by reciprocal transformations which can sometimes be used to generate new solutions from old \cite{Kingston}. A reciprocal transformation may be viewed as a map of the graph of a solution $u(x,t)$ of
\[
u_t=(D(u)u_x)_x
\]
to that of a solution $u'(x',t')$ of
\[
u'_{t'}=\left(D'(u')u'_{x'}\right)_{x'}.
\]
The map $(x,t,u)\to (x',t,u')$ is def\/ined by
\begin{gather}\label{reciprocal}
dx'=u\,dx+D(u)u_x\,dt,\qquad dt'=dt,\qquad u'=\frac{1}{u}.
\end{gather}
It can be shown that this induces the transformation of dif\/fusivities
\begin{gather}
D'(s)=\frac{1}{s^2}D\left(\frac{1}{s}\right).
\label{difftransform}
\end{gather}
The dif\/fusivity of interest here
\[
D(u)=\frac{1}{1+u^2}
\]
is seen to be invariant under the reciprocal transformation,
\[
D'(u')=\frac{1}{1+{u'}^2}=D(u').
\]
This raises the possibility of using the known solutions and generating new solutions of~(\ref{gendiff}) and hence new curve f\/lows by quadrature. However, it turns out that the reciprocal transformation of solutions of (\ref{gendiff}) acts geometrically trivially on solutions of the curve shortening equation.

\begin{proposition}\label{reciprocalThm}
Suppose $u(x,t)$ is a solution of \eqref{gendiff} and $y(x,t)$ the corresponding solution of the curve shortening equation. Let $(x',t',u')$ be the reciprocal transformation of $(x,t,u)$. Then the induced transformation on solutions of \eqref{CSE_intro} satisfies $(x',t',y')=(y,t,x+c)$, $c ={\rm const}$. That is, for each $t$, the reciprocal transformation of a known solution of \eqref{gendiff} induces a  reflection $($up to additive constant$)$ of the solution $y(x,t)$ in the line $y=x$.
\end{proposition}

\begin{proof}
Equation (\ref{gendiff}) and the reciprocal transformation (\ref{reciprocal}) permits us to write
\[
x'=\int_0^xu(x_1,t)dx_1-\int_a^tV_0(t_1)dt,
\]
where $a$ is a constant and $V_0(t)=-D(u(0,t))u_x(0,t)$. In terms of Cartesian coordinates $(x,y)$,
\begin{gather*}
x' =\int_0^xy_x(x_1,t)dx_1-\int_a^tV_0(t_1) =y(x,t)-y(0,t)+\int_a^tD(y_x(0,t_1))y_{xx}(0,t_1)dt_1\\
\phantom{x'}{} =y(x,t)-y(0,t)+\int_a^ty_t(0,t_1)dt_1=y(x,t)-y(0,a).
\end{gather*}
Then
\[
u'=1/u=1/y_x=x_y=x_{x'}.
\]
Now as $y'$ satisf\/ies  $y'_{x'}=u'$ just as $y_x=u$, then it follows that $y'=x+H(t')$. However, since $y'(x',t')$ is to satisfy the curve shortening equation
\[
y'_{t'}=\frac{y'_{x'x'}}{1+(y'_{x'})^2},
\]
it follows that $H={\rm const}$.
\end{proof}

Let $f_1$ be a function such that the ref\/lection of its graph def\/ines a function $f_2$. In gene\-ral~$f_1$ and~$f_2$ will satisfy dif\/ferent dif\/ferential equations and this can be very useful.
For example, the nonlinear dif\/fusion equation with dif\/fusivity $D(s)=s^{-2}$, seen from  (\ref{difftransform}) to be directly transformable to the classical linear heat equation $(k=1)$, is one of many integrable equations that can be linearized in its potential form by the hodograph transformation \cite{Clarkson}:  $(x',y')=(y,x)$ given in Proposition~\ref{reciprocalThm}. However, from a geometric point of view a curve and its ref\/lection are indistiguishable. Thus invariance of (\ref{gendiff}) under reciprocal transformations does not increase the class of known curve f\/lows. This highlights the dif\/f\/iculty of constructing solutions of the curve shortening equation and partly explains why so few interesting exact solutions are known.

\section{Functionally separable nonlinear heat equations}\label{section3}

Apart from symmetry methods, one can seek reductions that arise from second or higher order dif\/ferential constraints rather than the f\/irst order dif\/ferential constraints implied by classical Lie reduction. Unfortunately, there is no general procedure for seeking such reductions and much ef\/fort has gone into devising new reduction strategies for partial dif\/ferential equations. Thus in Doyle and Vassiliou~\cite{Doyle} the authors managed to classify all one-dimensional sourceless heat equations
\begin{gather}\label{NLheat}
u_t=(D(u)u_x)_x
\end{gather}
that admit separation of variables in {\it some} f\/ield variable $\bar u$. That is, one asks for a change of f\/ield variable $\bar{u}=m(u)$, such that the image of (\ref{NLheat}) under the change of variable, constrained by  the additively separable condition
\[
\bar{u}_{xt}=0,
\]
is a dif\/ferential system of f\/inite type that can therefore be solved by ordinary dif\/ferential equations. Indeed the resulting dif\/ferential system has the general form
\begin{gather}\label{NLheatSep}
\bar{u}_t=f(\bar{u})\bar{u}_{xx}+g(\bar{u})\bar{u}^2_x,\qquad \bar{u}_{xt}=0,
\end{gather}
for some functions $f>0$, $g$. It is proven in \cite{Doyle} that for any such pair $f$, $g$ there is a change of dependent variable that transforms $(\ref{NLheatSep})_1$ to (\ref{NLheat}). It turns out that system (\ref{NLheatSep}) has a maximal 3-parameter solution space for any given pair~$f$,~$g$.  In this manner the authors obtain exactly nine distinct dif\/fusivities $D$ up to the maximal transformation group that preserves the canonical form (\ref{NLheat}) for which the maximal 3-parameter solution space is achieved. In many cases the 3-parameter solution was constructed. This considerably extends the list of nonlinear dif\/fusion equations for which explicit solutions are available. One equation on the Doyle--Vassiliou list is the nonlinear heat equation
\begin{gather}\label{DVeq}
u_t=\left(\frac{u_x}{1+u^2}\right)_x.
\end{gather}
Thus the dif\/ferential 1-form
\[
\omega=u\, dx+\frac{u_x}{1+u^2}\,dt
\]
is closed on solutions of (\ref{DVeq}). It is easy to see that any function $y(x,t)$ satisfying $dy=\omega$ is a~solution of the curve shortening equation
(\ref{CSE_intro}). The solutions of (\ref{DVeq}) constructed in \cite{Doyle} are the functions
\[
u(x,t)=U\big(\sigma(x+a),\sigma^2(t+b)\big),
\]
where $a$, $b$ and $\sigma\neq 0$ are arbitrary constants and $U$ is one of the functions
\begin{gather}
 U(x,t)=\tan x,\nonumber\\
U(x,t)=\frac{x}{\sqrt{-x^2-2t}},\nonumber\\
U(x,t)=\frac{\pm 1}{  \sqrt{e^{2(x-t)}-1}},\nonumber\\
U(x,t)=\frac{\sinh x}{\sqrt{-\cosh^2x-e^{-2t}}},\label{DVsols}\\
U(x,t)=\frac{\pm\cosh x}{\sqrt{-\sinh^2x+e^{-2t}}},\nonumber\\
U(x,t)=\frac{\sin x}{\sqrt{\cos^2x-e^{2t}}},\nonumber\\
U(x,t)=\frac{\sin x}{\sqrt{\cos^2x+e^{2t}}}.\nonumber
\end{gather}
Each solution (\ref{DVsols}) provides a 3-parameter family of explicit curve shortening f\/lows except for the complex-valued, $\text{(\ref{DVsols})}_4$.

The function $y(x,t$) may be obtained from $U(x,t)$ simply by integrating in $x$, then adding a~suitable function of $t$.
\begin{enumerate}\itemsep=0pt
\item $\text{(\ref{DVsols})}_1$ then leads to the Calabi ``grim reaper'' travelling wave.
\item  $\text{(\ref{DVsols})}_2$ integrates to the well-known shrinking circle
$x^2+y^2=2(t_0-t)$.
\item Integration of $\text{(\ref{DVsols})}_3$ merely produces a horizontal version of the vertical grim reaper.
\item $\text{(\ref{DVsols})}_4$ is complex-valued, not considered further in the current practical context.
\item $\text{(\ref{DVsols})}_5$ is equivalent to  $\text{(\ref{DVsols})}_7$ by a rotation in the $xy$-plane.
\item The f\/inal two solutions are related to those previously presented by King~\cite{King}, expressed in the time-reversed form as examples of f\/inger growth; these deserve closer inspection.
\end{enumerate}

\subsection{Exact heat f\/low of a convex curve}

The studies of Gage--Hamilton \cite{GageHamilton} and Grayson \cite{Grayson1} on f\/low by curvature of embedded plane curves is a justly celebrated chapter in dif\/ferential geometry.

\begin{theorem}[Gage--Hamilton]\label{GageHamiltonThm}
Let $\gamma_0: I_x\to \mathbb{R}^2$ be a convex curve embedded in the plane. Let~$\gamma_0$ evolve by the curve shortening flow. That is,
\begin{gather*}
\frac{\partial \gamma}{\partial t}(x,t)=\kappa(x,t)\mathbf{n}(x,t),\qquad
\gamma(x,0)=\gamma_0(x),
\end{gather*}
where $\kappa(x,t)$ is the Euclidean curvature of $\gamma(x,t)$. Then the curve remains convex and becomes circular as it shrinks in the sense that
\begin{enumerate}\itemsep=0pt
\item[$1)$] the ratio of the inscribed radius to the circumscribed radius approaches~$1$;
\item[$2)$] the ratio of the maximum to the minimum curvature approaches~$1$;
\item[$3)$] the higher order derivatives of the curvature converge to zero uniformly.
\end{enumerate}
\end{theorem}

$\text{(\ref{DVsols})}_6$  provides the only known example of an explicit, non-self similar curve shortening f\/low in case the initial curve $\gamma_0$ is a closed, convex embedded plane curve which is not a circle. We use results described in Section~\ref{section2}.
The sixth  function in (\ref{DVsols}) is
\begin{gather}\label{solDV6}
u=\frac{\sin x}{\sqrt{\cos^2 x-e^{2t}}}.
\end{gather}

In the case of (\ref{solDV6}), the corresponding solution of (\ref{CSE_intro}) is
\begin{gather}\label{CSEsolDV6}
y=t-\ln \left(\cos x+\sqrt{\cos^2x-e^{2t}}\right).
\end{gather}
Clearly if $y$ is a solution of (\ref{CSE_intro}) then so is $-y$. The two solutions join smoothly along $y=0$ and can be jointly expressed in the simple implicit form
\begin{gather}\label{CSEsolDV6_new}
\cosh y-e^{-t}\cos x=0.
\end{gather}
This solution is recorded in \cite{King} and shown here to arise from functional separation. Solutions of this type are also studied in \cite{Daskalopoulos} and referred to as `Angenent ovals', although explicit solutions are not written down in the latter paper. We will now study some properties of this solution.
For each $t\in(-\infty,0)$ equation (\ref{CSEsolDV6_new}) def\/ines a closed, convex ``oval-shaped'' curve which is symmetric about the $x$- and $y$-axes  for
\[
x\in \left(-\cos^{-1}\left(e^{t}\right),  \cos^{-1}\left(e^{t}\right)\right),\qquad
y\in \left(-\cosh^{-1}\left(e^{-t}\right),  \cosh^{-1}\left(e^{-t}\right)\right).
\]
By analogy with an ellipse, eccentricity may be def\/ined as
\begin{gather*}
\epsilon(t)= \sqrt{1-(x_{\max}/y_{\max})^2}
= \sqrt{1-\left[\frac{\cos^{-1}\left(e^{t}\right)}{\cosh^{-1}\left(e^{-t}\right)}\right]^2}
= \sqrt{\frac{2}{3}|t|}+O\left(|t|^3\right),
\end{gather*}
showing approach to circularity $(\epsilon \to 0)$ as $t$ approaches the extinction time 0.
The curvature at each point $(x,t)$ of the curve is
\[
\kappa(x,t)=\frac{e^{-t}\cos x}{\sqrt{e^{-2t} -1}}.
\]
For each $t\in\left(-\infty,0\right)$, the maximum curvature occurs at $x=0$ with value $\kappa_{\max}=e^{-t}/\sqrt{e^{-2t}-1}$, while the minimum occurs at the extremities along the minor axis, $x=\pm \cos^{-1} \left(e^t\right)$ with value $\kappa_{\min}=1/\sqrt{e^{-2t}-1}$. Hence
\[
\lim_{t\to 0^-}\frac{\kappa_{\max}}{\kappa_{\min}}=\lim_{t\to 0^-} e^{-t}=1,
\]
verifying the Gage--Hamilton theorem and more specif\/ically showing that the ratio of maximum to minimum curvature is an exponential function converging to unity.  Of course, the curvature itself is an unbounded function of time as the f\/low continues toward extinction.

It transpires that for this curve f\/low, arclength can be expressed as a function of time in terms of the incomplete elliptic integral of the f\/irst kind.

\begin{figure}[t]
\centering
\includegraphics[width=9cm]{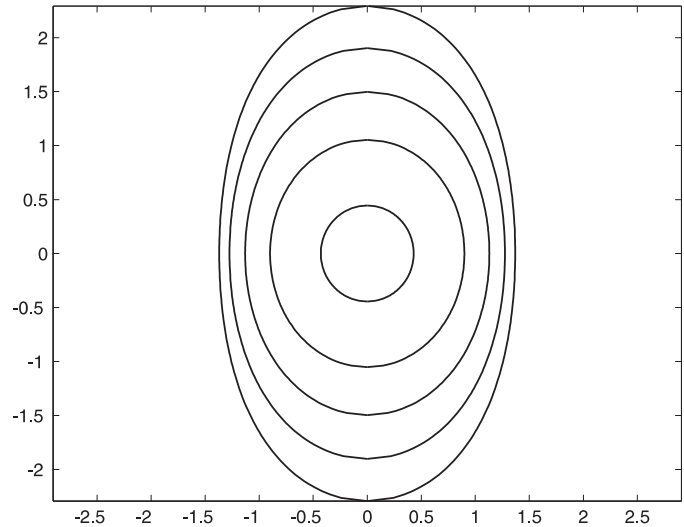}
\caption{ Evolution by heat shrinking f\/low of the curve def\/ined by $\cosh y-5\cos x=0.$}
\label{fig0}
\end{figure}

At very early times $t\ll 0$, the solution with $x\in (-\pi/2,\pi/2)$ is asymptotic to two grim reapers joined smoothly and approaching each other with constant speed:
\begin{gather*}
y=\pm(t-\log \cos (x)-\log  2) + O\big(e^{4t}\sec^4(x)\big).
\end{gather*}

A question that may be posed is: why construct curve shortening f\/lows by f\/irst solving (\ref{DVeq}) rather than solving (\ref{CSE_intro}) directly? Remarkably, it can be shown that
\begin{proposition}\label{SeparableCSEthm}
The image of the curve shortening equation \eqref{CSE_intro} under the transformation $y\mapsto m(y)$
does not have a $($maximal$)$ $3$-parameter family  of joint solutions with the linear wave equation $y_{xt}=0$ in any field variable $m(y)$ except $m=\text{identity}$ in which case the solution gives rise to the grim-reaper flow.
\end{proposition}

\begin{proof} Construct the image of the curve shortening equation (\ref{CSE_intro}) under the change of variable $y\mapsto m(y)$ and apply the Cartan--K\"ahler theorem to the dif\/ferential system consisting of the transformed curve shortening equation and the constraint $y_{xt}=0$.
\end{proof}

This is in sharp contrast to the rich separability properties of nonlinear heat equations (\ref{NLheat}) as discussed in Section~\ref{section2}. Thus starting with (\ref{DVeq}) appears to be an important f\/irst step in constructing non-trivial curve f\/lows\footnote{In view of the relationship between $u(x,t)$ satisfying (\ref{DVeq}) and $y(x,t)$ satisfying (\ref{CSE_intro}), one might try for the higher order constraint $y_{xxt}=0$ rather than $y_{xt}=0$. This is possible but the calculations rapidly become very complicated as the order increases.}.

\subsection{Decaying periodic solution initially close to square wave}

 For a curve f\/ixed at two end-points, we prescribe the boundary conditions
 \begin{gather*}
 y=0, \qquad x=0,\ell.
 \end{gather*}
 We now def\/ine dimensionless space and time variables
 $X=x/\ell$, $Y=y/\ell$, $\tau=Bt/\ell^2$,
in terms of which the curve shortening equation\footnote{Also known in the materials science community as the {\it Mullins equation}~\cite{Mullins}.} is
 \begin{gather}
 Y_\tau = \frac{Y_{XX}}{1+Y_X^2},
 \label{Mullins2b}
 \end{gather}
 to be solved subject to boundary conditions
  \begin{gather}
 Y = 0, \qquad X=0,1
 \label{bcs2}
 \end{gather}
 and continuous initial conditions $Y(X,0)=Y_0(X).$

Integrating the seventh member of the list (\ref{DVsols})  and then applying translational and scaling invariance  transformations, we obtain
 \begin{gather}
 Y=\frac{1}{K} \ln\left(\frac{\sqrt{\exp(2K^2[\tau-\tau_0])+\cos^2(K[X-X_0])}+\cos(K[X-X_0])}{\exp(K^2[\tau-\tau_0])}\right),
 \label{gensoln}
 \end{gather}
 with $K$, $X_0$, $\tau_0$ arbitrary constants.
 It may be verif\/ied that (\ref{gensoln}) satisf\/ies (\ref{Mullins2b}). In fact, when $\tau-\tau_0$ is large and negative,
 \begin{gather*}
 Y=\frac{\ln 2}{K}-K[\tau-\tau_0]+\frac{1}{K}  \ln \cos (K[X-X_0]) +O\left(\frac{e^{-2K^2[\tau_0-\tau]}}{\cos(K[X-X_0])}\right).
 \end{gather*}
 This shows that away from the singularities of   $\ln\cos(X-X_0)$, the solution is asymptotic in the distant past to the Calabi ``grim reaper'' solution of the curve-shortening f\/low \cite{Grayson1}.
 At all times, the solution~(\ref{gensoln}) is a  deformation of the grim reaper solution but now it is extended smoothly and periodically, without singularities over a domain of any length. Unlike in the grim reaper solution, there are f\/ixed points so that we may apply the Dirichlet boundary conditions~(\ref{bcs2}), which lead to
 \begin{gather}
 Y=\pm\frac{1}{K} \ln\left(\frac{\sqrt{\exp(2K^2[\tau-\tau_0])+\sin^2(K X)}+\sin(K X)}{\exp(K^2[\tau-\tau_0])}\right),
 \label{newsoln}
 \end{gather}
 with
 \[
 K=\frac{n\pi}{\ell}.
 \]
The amplitude of $Y(X,\tau)$, which is the value of $|Y(X,\tau)|$ at $X=\pm\pi/(2K)$, is approximated by
 \begin{gather*}
 Y_{\max}=K  [\tau_0-\tau]+(\ln 2)/K +O\left(\exp\left(-K^2|\tau-\tau_0|\right)\right).
 \end{gather*}
 In terms of dimensional quantities,
 \begin{gather}
 y_{\max}=\frac{n\pi B[t_0-t]}{\ell}+\ell\frac{\ln 2}{n\pi}+O\left(\exp\left(-n^2\pi^2B[t_0-t]/\ell^2\right)\right),
  \label{lowbound}
  \end{gather}
  where $y_{\max}=\ell Y_{\max}$ and $t_0=\ell^2 \tau_0/B$.
 This shows that at early times, the amplitude decreases linearly as a function of time.

\begin{figure}[t]
\centering
\includegraphics[width=9cm]{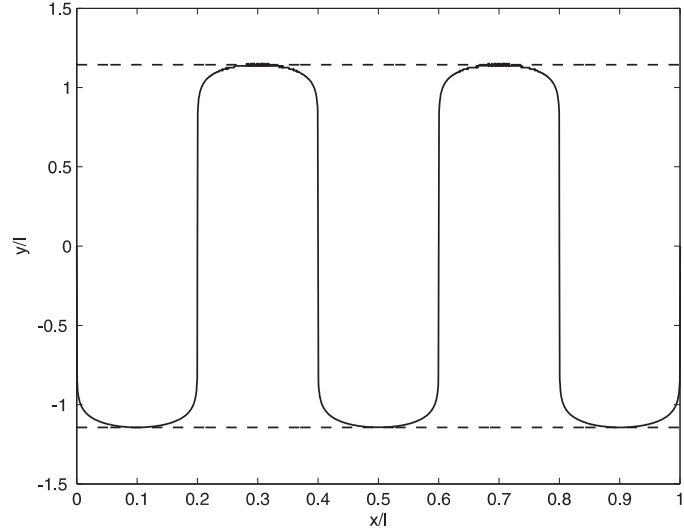}
\caption{Filled curve: exact solution with $\tau-\tau_0=-0.07$, $K=5\pi$. Dashed lines: approximate bounds $y=\pm K [\tau_0-\tau]+(\ln 2)/K$.}
\label{fig3}
\end{figure}

\begin{figure}[t]
\centering
\includegraphics[width=9cm]{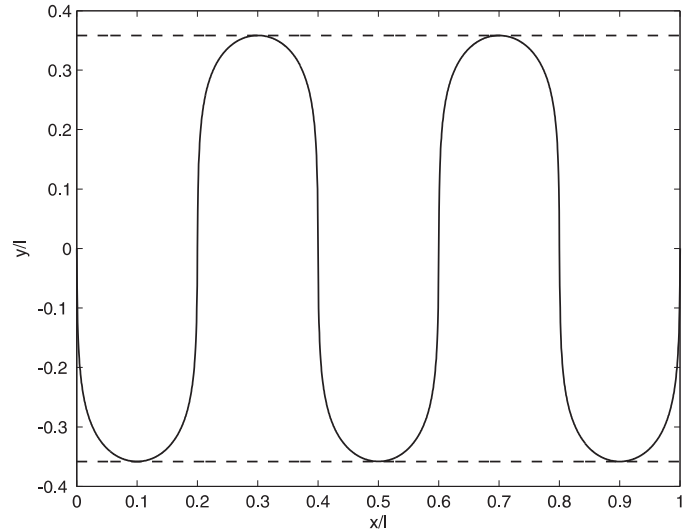}
\caption{Filled curve: exact solution with $\tau-\tau_0=-0.02$, $K=5\pi$. Dashed lines: approximate bounds $y=\pm K [\tau_0-\tau]+(\ln 2)/K$.}
\label{fig4}
\end{figure}

 As an example, the solution is graphed for the case $n=5$ and with the minus sign preceding the right hand side of (\ref{newsoln}). The solution at early times is shown in Figs.~\ref{fig3},~\ref{fig4}. Although for large $\tau_0$, the initial condition  resembles a periodic square wave, it actually converges pointwise to a dif\/ferentiable  bounded and periodic grim reaper as $\tau_0$ approaches $\infty$.

When $K^2[\tau-\tau_0]$ is large and positive, the solution is smoother, approximated by a single sine wave,
\begin{gather}
Y=\pm\frac{\exp(-K^2[\tau-\tau_0])}{K}\sin(K x)+O\left(\exp(-2K^2[\tau-\tau_0])\right),
\label{highbound}
\end{gather}
whose amplitude  decreases exponentially in time. This is the sinusoidal solution of the classical linear dif\/fusion equation that approximates the curvature driven dif\/fusion  equation at large times in the small-slope approximation.

	The number of extrema $n$ within the f\/ixed domain $[0,1]$ may be freely chosen. For large values of the parameter $K^2\tau_0$, as in Fig.~\ref{fig3}, the initial condition is described well as a periodic square wave, resembling a dif\/fraction grating. The time scale for decay may be viewed as the time at which the formal expression~(\ref{lowbound}) for $Y_{\max}$ is zero. This is
\[
t=t_0+\frac{\ell^2\ln 2}{n^2\pi^2B}.
\]
		
	Fig.~\ref{fig3} evidences  considerable smoothing of the initial conditions but at this time, the solution does not yet resemble a simple sinusoid. At larger times, the solution is close to a single sinusoidal wave. In the regime of the sinusoidal prof\/ile, from~(\ref{highbound})  decay times are shorter, of the order of the standard time $(\ell/n\pi)^2/B$ that is familiar from exponential decay of Fourier modes in linear dif\/fusion models (e.g.~\cite{Carslaw}) when the half-wavelength $\ell/n$ is the typical distance between neighboring regions of high and low mass concentration.

\section{Surface evolution by evaporation and condensation}\label{section4}

For some metals such as gold, surface dif\/fusion persists for several thousand years as the dominant surface transport mechanism but for others such as magnesium, after a time less than a day, surface evolution occurs predominantly  by evaporation-condensation. As described by Mullins~\cite{Mullins}, the Gibbs--Thomson relation for evaporation rate in terms of def\/icit from equilibrium pressure over a curved surface, leads to a second-order equation for dif\/fusion by mean curvature. In terms of two-dimensional  Cartesian coordinates~$(x,y)$ and time~$t$, the Mullins equation for points on a material surface is
\begin{gather*}
y_t = B\frac{y_{xx}}{1+y_x^2},
\end{gather*}
where $B $ is  constant.  This equation applies to two dimensional cross sections of solids when surface nano-scale features such as grooves, ridges and furrows extend rectilinearly into the third dimension. Because of the nonlinearity, very few useful exact solutions to this equation are known~\cite{Taylor}, except in a linear approximation. Two decades ago~\cite{Broadbridge}, the exact solution was constructed in parametric integral form for nonlinear surface evolution near a  symmetric grain boundary, with constant slope at the grain boundary groove, initial f\/latness and zero displacement at inf\/inity. A similar  procedure produces the more general ``open angle'' solutions for deposition in a wedge~\cite{Ishimura}. Subsequently, the fourth-order Mullins equation for curvature-driven surface dif\/fusion on an almost-isotropic material, was solved with boundary conditions representing a symmetric grain boundary~\cite{PBPT,PTPB,PBJG}.

\looseness=-1
In the context of surface evaporation, the parallel asymptotes of the grim reaper solution represent a long thin metallic foil that is evaporating at the ends. The constant travelling-wave speed of the grim reaper solution shows that evaporation will take place at a constant rate that depends on foil thickness as well as the evaporation coef\/f\/icient $B$. The thickness $\Delta x$ is the distance between the two asymptotes, $\Delta x=\pi B/c$. Hence the steady evaporation rate at the end of a strip of metallic foil will be $c=\pi B/\Delta x$. For example, for a foil of a few microns in thickness made of the volatile metal~Mg, this rate will be of the order of one millimetre per millenium.

In the context of metal surface smoothing, the periodic solution of the previous section predicts the smoothing of initial conditions that resemble a dif\/fraction grating.
 For a surface f\/ixed at two end-points, we prescribe the boundary conditions
 \begin{gather*}
 y = 0, \qquad x=0,\ell.
 \end{gather*}
 Physically, this corresponds to the surface being clamped and shielded from the surrounding atmosphere outside of the exposed spatial domain $[0,\ell]$.

 \subsection{Evaporation from anisotropic crystals}
C.~Herring \cite{Herring}, showed that for anisotropic crystals, surface energy is proportional to $b(\phi)(\gamma +\gamma ''(\phi) )$, where $\phi$ is the polar angle $\arctan y_x$, $\gamma$ is surface tension and $b$ is an evaporation coef\/f\/icient. An equation of the form
	\begin{gather}
	y_t=D(y_x)y_{xx},
	\label{AnisDiff}
	\end{gather}
	implying the standard nonlinear dif\/fusion equation (\ref{NLheat}) with $u=y_x$, may be regarded as an anisotropic form of the Mullins evaporation-condensation equation
	\begin{gather*}
	y_t=B(y_x)\frac{y_{xx}}{1+y_x^2},
	\end{gather*}
	with surface slope-dependent anisotropy factor $B(y_x)=D(y_x)(1+y_x^2)$, which originates in the physical derivation from a constant multiple of $b(\phi)(\gamma+\gamma ''(\phi))$ (e.g.~\cite{Tritscher97}). Crystalline materials are indeed anisotropic, with the evaporation coef\/f\/icient minimized when the cut surface is aligned with crystal planes. The  group of equivalence transformations of this class of equations includes the general linear group $GL(2,{\mathbb R})$. By the polar decomposition theorem, each invertible linear transformation can be decomposed as an othogonal transformation followed by multiplication by a~positive def\/inite symmetric dilatation matrix. Under a rotation about the origin by angle~$\alpha$,
	\begin{gather*}
	\bar x=x \cos \alpha-y \sin\alpha,\qquad \bar y=x \sin \alpha+y \cos\alpha,\qquad \bar t=t,\\
	\bar u=\bar y_{\bar x}=\frac{u+\tan \alpha}{1-u \tan \alpha},\qquad
	y_x=\frac{-\sin \alpha+\bar y_{\bar x}\cos \alpha}{\cos \alpha+\bar y_{\bar x}\sin \alpha},\qquad
	y_{xx}=\frac{\bar y_{\bar x\bar x}}{(\cos \alpha+\bar y_{\bar x}\sin \alpha)^3} .
 \end{gather*}
	The axis $\bar x=0$ has been rotated by angle $-\alpha/2$. Writing $\bar\theta=\arctan\bar y_{\bar x}$,
	\begin{gather*}
	y_t=\frac {{\mathbf n} \cdot{\mathbf e}_{\bar y}} {{\mathbf n} \cdot {\mathbf e}_y} \bar y_{\bar t}
=\frac{\cos \bar\theta}{\cos (\bar\theta-\alpha)} \bar y_{\bar t}
 = \frac{1}{\cos \alpha} \frac{1}{1+\bar y_{\bar x}\tan \alpha }  \bar y_{\bar t}.
	\end{gather*}
	Hence, by rotation, the isotropic nonlinear dif\/fusion equation (\ref{AnisDiff}) is equivalent to
	\begin{gather*}
	\bar y_{\bar t}=\bar D(\bar y_{\bar x})\bar y_{\bar x \bar x}, \qquad
	\bar D(s)=\frac{1}{(s \sin\alpha+\cos\alpha)^2} D\left( \frac{s \cos\alpha-\sin\alpha}{s \sin\alpha+\cos\alpha}\right).
	\end{gather*}
	Tritscher \cite{Tritscher} used this device of rotational equivalence classes to solve integrable forms of fourth-order surface dif\/fusion equations.
	In particular, after rotation by angle $\alpha=\pi/2$,
	\[
	\bar D(\bar y_{\bar x} )=\frac{1}{\bar y_{\bar x}^2} D\left(\frac{-1}{\bar y_{\bar x}}\right).
	\]
	By following the $\pi/2$ rotation by a trivial ref\/lection $\bar x \to-\bar x$, we recover the result of the reciprocal transformation (\ref{difftransform}).

	By the principal axis theorem, a positive symmetric matrix can be written  as $O^tQO$, where $O$ is a rotation matrix and $Q$ is diagonal, $Q_i^j=a_i\delta_i^j$ with $a_j>0$. Therefore, to consider the ef\/fect of an additional dilatation, we need only consider the ef\/fect of a diagonal rescaling:
	 \begin{gather*}
	 \bar x=a_1x,\qquad \bar y=a_2y,\qquad
	 \bar y_t=a_1^2D\left(\frac{a_1}{a_2}\bar y_{\bar x}\right)\bar y_{\bar x \bar x},
	 \end{gather*}
	 which although trivial, allows us to construct simple anisotropic models from the rotationally invariant isotropic model.
	
	\section{Anisotropic analogs of isotropic model solutions}\label{section5}
	
	The scale invariance group still applies to the general anisotropic dif\/fusion equation (\ref{AnisDiff}). Therefore both expanding and shrinking types of similarity solution exist  but they may be signif\/icantly dif\/ferent from those of the isotropic model. The grain boundary groove solution still exists, as can be seen from the solvability of the reduced boundary value problem on ${\mathbb R}^+ \times {\mathbb R} ^+$,
	\begin{gather*}
	u=F(\rho),\qquad \rho=xt^{-1/2},\\
		\frac{-\rho}{2}F'(\rho)=\frac{d}{d\rho}[D(F)F'(\rho)],\\
		F(0)=m>0,\qquad
		F(\rho)\to 0, \qquad \rho\to\infty.
		\end{gather*}
		
		The grain-boundary groove solutions for all models have some common features. Although it represents a highly anisotropic material, the linear model groove solution approximates that of an isotropic model for groove slopes of up to~0.5, which was used in Mullins' original paper~\cite{Mullins}. However, whereas the linear model predicts that the groove depth increases in proportion to $m$,  the groove depth increases very slowly, of order  $(\log m)^{0.5}$ at large $m$ for the isotropic model~\cite{Karciga}.

\subsection{Anisotropic homothetically shrinking closed curve}
	
	For anisotropic evaporating materials, the closed-curve homothetic solution represents the f\/ixed-shape cross section of an evaporating wire, which is circular when the material is isotropic.
	
	The homothetic evaporating closed-curve solution satisf\/ies
		\begin{gather*}
		y=[t_0-t]^{1/2}G(\rho),\qquad \rho=x[t_0-t]^{-1/2},\\
		\frac{-G}{2}+\frac{\rho}{2}G'(\rho)=G''(\rho)D(G'),\\
		G'(0)=0,\qquad
		G(\rho_0)=0,\qquad
		G'(\rho)\to \infty, \qquad \rho\to\rho_0.
		\end{gather*}
		
	This implies
		\begin{gather}
		u=F(\rho), \qquad F=G',\nonumber\\
		\frac{\rho}{2}F'(\rho)-D'(F)(F')^2=F''(\rho)D(F),\label{reduced} \\
				F(0)=0,\qquad
		F(\rho)\to \infty, \qquad \rho\to\rho_0.\nonumber
		\end{gather}
		
		For some functions $D(u)$, the homothetic closed-curve solution does not exist. For example, with  the linear model with constant $D$, the general solution satisfying $F(0)=0$ must  be
\[
F(\rho)=\frac{\sqrt\pi}{2}A \, \text{erf}\,(\rho)=A\int_0^\rho e^{-s^2/4}ds ,\qquad A\in{\mathbb R}.
\]
		Since this cannot take an inf\/inite value at any point $\rho=\rho_0\in {\mathbb R}$, the closed-curve homothetic solution does not exist.
		
		 Let us now assume that
		$F(\rho)\approx A_0(\rho_0-\rho)^\nu$. If a homothetic closed-curve solution exists,
		$G(\rho_0)\in(-\infty,0)$, and
\[
0=G(\rho_0)=G(0)+\int_0^{\rho_0}F(s)ds.
\]
		This can hold only if the integral does not diverge, implying
		\begin{gather*}
		-1<\nu<0 .
		\end{gather*}
		 Now we suppose that at large $u, D(u)$ is asymptotic to a power law $D(u)\approx D_0 u^n$. Then by balancing terms at the leading order in $\rho_0-\rho$, (\ref{reduced}) implies
		\begin{enumerate}\itemsep=0pt
		\item[$(i)$] $n=1/\nu$, which is less than $-1$, and  $\rho_0=-2\nu D_0A_0^{1/\nu}$,  or,
		\item[$(ii)$] $n=\frac{1-\nu}{\nu}$, which is less than $-2$.
		\end{enumerate}

		 For the linear model, the anisotropy factor  is $B(u)=(1+u^2)D$ which diverges when the curve is vertical. In the current application, we are interested in cases of realistic anisotropy for which the evaporation coef\/f\/icient and the anisotropy factor are bounded, and  the latter with a~minimum value greater than zero:
\[
\forall \, u\in{\mathbb R},\qquad 0<B_{0}<B(u)<B_{\infty}<\infty .
\]
	 That statement must be true, independent of the orientation of the coordinate axes. For example after rotation by~$\pi/2$ it must be true that $\bar D(\bar u)\to \bar D_0\in(0,\infty)$ as $\bar u\to 0$. By rotating back to the original orientation, this implies
	 \begin{gather*}
	 D(u)=u^{-2}\bar D(u)\sim \bar D_0 u^{-2}.
	 \end{gather*}
	 This physical restriction rules out scenario $(ii)$ in the above leading-order analysis.
	 The simplest example satisfying the restriction is $D(u)=1/(1+(\beta u)^2)$ ($\beta\in{\mathbb R}$  constant) for which the anisotropy factor $B(y_x)$ varies between~1 and~$\beta^2$. In this case the homothetic solution is simply an ellipse
\[
\frac{x^2}{2[t_0-t]}+\frac{y^2}{2\beta^2[t_0-t]}=1
\]
	elongated in the direction of weakest evaporation.

	\subsection{Grim reaper solution for anisotropic material}
	
	The grim reaper solution is a travelling wave  constrained between two vertical asymptotes. This implies a steady state solution of the equation $u_t=[D(u)u_x]_x$.  For arbitrary $D(u)$, there exists  a two-parameter steady state solution
	\[
	u_s(x)=K^{-1}(cx+c_2),
	\]
where	
	\[
	K(u)=\int_0^uD(s)ds
	\]
	which is an increasing  invertible function because $D(s)>0$. 	This integrates formally to
	\[
	y(x,t)=y_0(t)+\int_0^xu_s(x_1,t)dx_1.
	\]
	It then follows from (\ref{AnisDiff}) that $y_0(t)$ can only be a constant-velocity translation, $y_0=c(t-t_0)$. This travelling wave solution does not necessarily have a vertical asymptote. For example, there is no such asymptote when $D$ is constant. However if $D(s)$ satisf\/ies the physical requirement $D(s)\approx s^{-2}$, it follows that $K(u)$ has a f\/inite limit as $u\to\infty$, therefore $K^{-1}$ has a vertical asymptote at some location $x=x_0$, duplicated at $x=-x_0$ if $D(s)$ is an even function, as is commonly the case when the $x$-axis denotes the orientation of the crystal planes from where evaporation is weakest.

\section{Anisotropic models allowing functional separation}	\label{section6}
	
	The classif\/ication of Doyle and Vassiliou gives all functions $D(s)$ for which functional separation of variables is possible in the general form
	\begin{gather}
	\bar u=f(u)=v(x)+w(t),
	\label{Separated}
	\end{gather}
	 with $f$ an invertible function.
	
	 The simplest example of a physically feasible isotropic model is simply that obtained from the isotropic dif\/fusion equation by unequally rescaling $x$ and $y$. For example, Fig.~\ref{fig0} could  be dilated in one direction, displaying a non-homothetic closed curve approaching a homothetically shrinking ellipse.

	The only other member of the Doyle--Vassiliou list with realistic anisotropy is much more complicated:
	\begin{gather}
	D(u)=D_0\cos(z(Au)), \qquad
	Au=\int_0^{z}(\cos  s)^{-3/2}ds, \qquad -\pi/2<z<\pi/2.
	\label{kcos}
	\end{gather}
	
\begin{figure}[t]
\centering
\includegraphics[width=9cm]{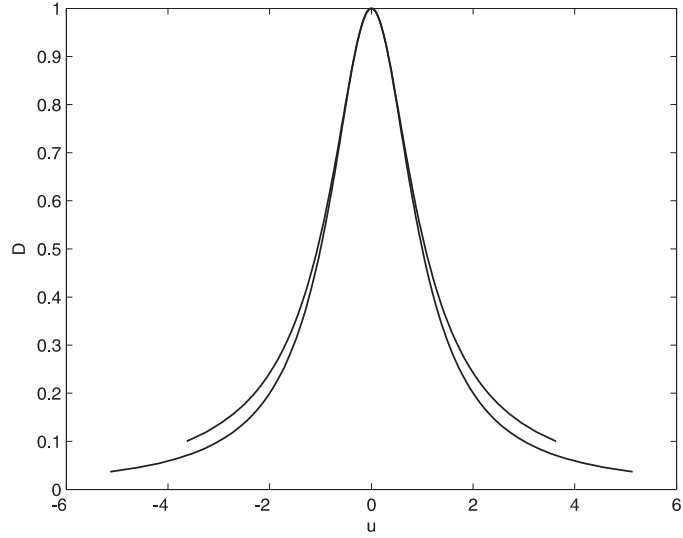}
\caption{Nonlinear dif\/fusivity function $D(u)$ for isotropic evaporation (lower) and anisotropic model (upper).}
\label{Kplot_1}
\end{figure}

	This model is  close to isotropic when $A=\sqrt 2$. In that case, it is easy to show that when $u(=y_x)$ is small, $B(u)=1+O(u^4)$ and $B\to2$ as $u\to\infty$. In Fig.~\ref{Kplot_1}, the function $D(u)/D_0$ for this weakly anisotropic model is compared to $D(u)/D_0$ of the isotropic model.

	For the sake of completeness, we construct the special solution $u(x,t)$ compatible with~(\ref{Separated}), that was not given explicitly by Doyle and Vassiliou~\cite{Doyle}. The parameter $A$ may be changed by rescaling $u$. For convenience, without loss of generality we now set~$A$ to~1. Also we may set~$D_0$ to~1 by using~$D_0t$ as the time coordinate. From the general approach of Doyle and Vassiliou~\cite{Doyle}, $\bar u$ is a sum of separated functions $v(x)$ and $w(t)$ satisfying
	\begin{gather}
	v'''(x)+\frac{1}{2} [v'(x)]^3=0, \label{veq}\\
	w'(t)=W(w)=\cos w\left[v''\cos v+\frac{1}{2}(v')^2\sin v\right] +\sin w\left[-v''\sin v+\frac{1}{2}(v')^2\cos v\right].\nonumber
	\end{gather}
	
	By construction, the following are f\/irst integrals of~(\ref{veq}), which can be verif\/ied by substitution:
	\begin{gather*}
	\lambda=v''(x)\cos v+\frac{1}{2}(v')^2\sin v,\qquad
	\mu=-v''(x)\sin v+\frac{1}{2}(v')^2\cos v.
	\end{gather*}
	It then follows that $w'(t)=\lambda \cos w+\mu \sin w$, and  that
	\begin{gather}
	t=\frac{2}{R^2}\int\frac{dw}{\sin(w+\delta)}, \qquad
	x=\frac{1}{R}\int_0^v\frac{ds}{\sqrt{\cos(s-\delta)}},
	\label{xsol}
	\end{gather}
	with  constants $x_0$, $t_0$, $R=\pm 2^{1/2}(\mu^2+\lambda^2)^{1/4}$ and $\delta=\arctan(\lambda/\mu)$, whose arbitrariness is of little consequence.
	Let $v'(x)=W(v)$, so $v''(x)=WW'(v)$. Then
	\begin{gather*}
	WW'(v)\cos v+\frac{1}{2}W^2\sin v=\lambda, \qquad
	-WW'(v)\sin v+\frac{1}{2}W^2\cos v=\mu\\
	\Rightarrow \quad -\lambda\tan v+\frac{1}{2}W^2\frac{\sin^2v}{\cos v}+\frac{1}{2}W^2\cos v=\mu \\
	\Rightarrow \quad W^2=2(\mu \cos v+\lambda \sin v)=R^2\cos(v-\delta)\quad
	\Rightarrow \quad v'(x)=\pm R\sqrt{\cos(v-\delta)},\\
	x=\frac{1}{R}\int_0^v\frac{ds}{\sqrt{\cos(s-\delta)}}.
	\end{gather*}
	
	When we include the parameter $B_0$, the general solution for $v$ and $w$ in terms of elementary functions, the standard elliptic integral $F(\theta |\frac{1}{2})$ and the standard Jacobi elliptic function $\text{sn}(X |\frac{1}{2})$, is
	\begin{gather*}
	w=2\arctan e^{R^2B_0t/2}-\delta \qquad \mbox{and} \qquad
	v=2\arcsin\left(\frac{1}{\sqrt 2}\,\text{sn}\left(\frac{R[x-x_0]}{\sqrt 2}\Big|\frac{1}{2}\right)\right)+\delta,
\end{gather*}
where
	\begin{gather*}
x_0=\frac{\sqrt 2}{R}F\left(\arcsin\left(\sqrt 2\sin\frac{\delta}{2}\right)\Big|\frac{1}{2}\right).
	\end{gather*}
	Since $u$ is a function of $w(t)+v(x)$, we see that the choice of parameters $R$ and $\delta$ has little consequence on the form of the solution $u(x,t)$. A shift in the phase variable $\delta$ has the ef\/fect of a translation in~$x$ by $x_0(\delta,R)$. A change of amplitude $R$ has the ef\/fect of rescaling~$x$ and~$t$ to~$Rx$ and $R^2t$.

\begin{figure}[t]
\centering
\includegraphics[width=9cm]{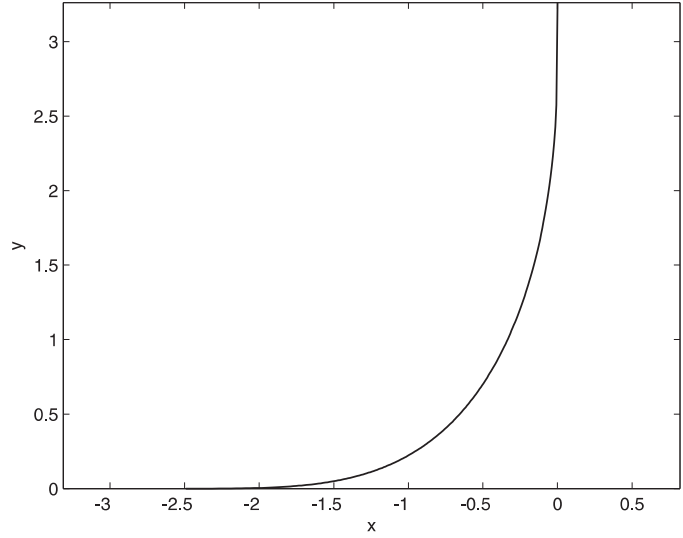}
\caption{Portion of solution curve for anisotropic model with $R=1$ and $\delta=0$, at $t=0$.}
\label{fig5}
\end{figure}

\begin{figure}[t]
\centering
\includegraphics[width=9cm]{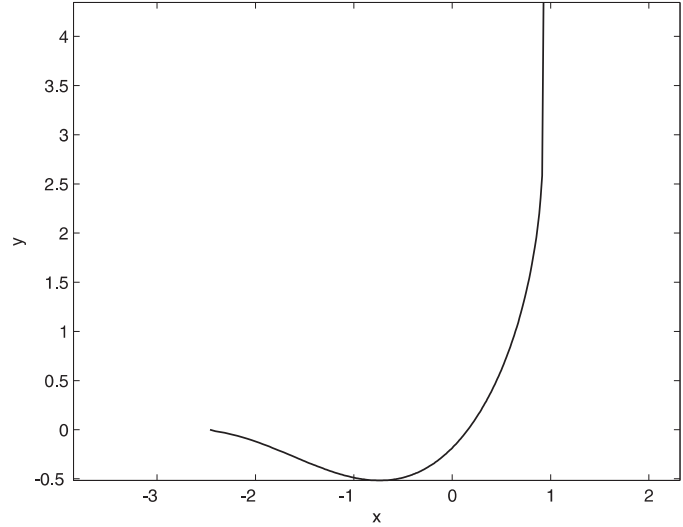}
\caption{Portion of solution curve for anisotropic model at $t=-2.0$.}
\label{fig6}
\end{figure}

\begin{figure}[t]
\centering
\includegraphics[width=9cm]{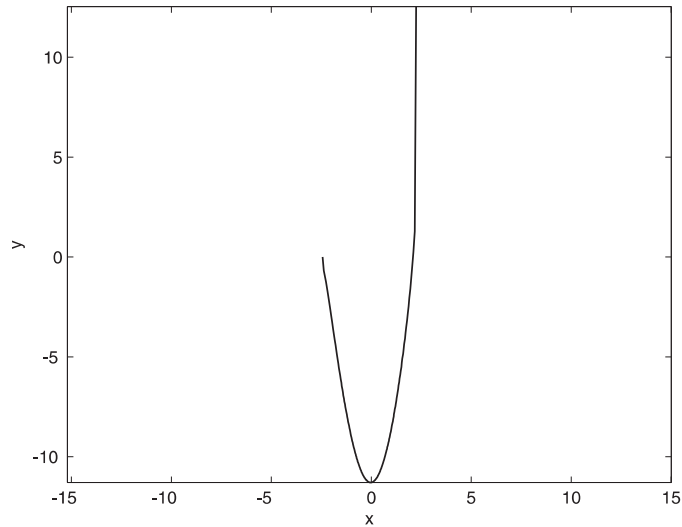}
\caption{Portion of solution curve for anisotropic model at $t=-8.0$.}
\label{fig7}
\end{figure}

	Modulo a time-dependent vertical translation, the curve $y(x,t)$ is obtained from $u(x,t)$ by integration. Since the integrands in (\ref{kcos}) and (\ref{xsol}) must be real valued, the constructed solution has a truncated domain. An example is given in Figs.~\ref{fig5}--\ref{fig7}.

As $t$ approaches $-\infty$, the solution approaches a steady state. Since $D(u)$ is symmetric, this steady state is a symmetric grim reaper. It is approximated in Fig.~\ref{fig7} by taking $t=-8.0$.

	The curve is not symmetric, as can be seen in Fig.~\ref{fig6}. This asymmetry causes its domain to shift slightly from right to left.
For $t\le 0$, the curve has a vertical asymptote at moving location
	\[
	x=x_u=\int_0^{\pi/2-w(t)}(\cos s)^{-1/2}ds
	\]
	and a local minimum at
	\[
	x=x_\ell=-\int_0^{w(t)}(\cos s)^{-1/2}ds.
	\]
		
\section{Conclusion}\label{section7}

Of the exact solutions to the curve shortening equation known to us, most can be obtained by Lie point symmetry reductions. The two interesting solutions that cannot be constructed in this way, can indeed be recovered by functional separation of variables for the standard nonlinear dif\/fusion equation (\ref{gendiff}) that is obtained from the curve shortening equation by dif\/ferentiation. The classif\/ication obtained by Doyle and Vassiliou \cite{Doyle} of nonlinear dif\/fusion equations that admit functional separation of variables, leads to the two  exact non-self-similar solutions with non-trivial initial conditions that appear to be achievable in an approximate sense in applications. In addition, it leads to a new separated solution for a physically realistic anisotropic evaporation-condensation dif\/fusion equation. Although the second order nonlinear surface evolution equations for slope $u(x,t)$ ($\equiv y_x(x,t)$ admit a number of possibilities for functional se\-pa\-ra\-tion of variables in Cartesian coordinates, we have proved that this is not possible for the equation~(\ref{CSE_intro}) governing $y(x,t)$, nor is it possible in a coordinate system consisting of canonical variables for a symmetry other than translation. This is in contrast with the point symmetry analysis, which leads to a richer array of possibilities for the evolution of $y(x,t)$ than for the evolution of $u(x,t)$.

The invariance of the isotropic equation (\ref{DVeq}) under the well known reciprocal transformation was shown (Proposition \ref{reciprocalThm}) to lead to no new planar curve heat f\/lows. The group of geometric equivalence transformations of the class of general anisotropic equations (\ref{AnisDiff}) includes not only the reciprocal transformation in the guise of a ref\/lection in the plane, but the whole general linear group. The equivalence is shown by carrying out the equivalence transformations explicitly. Physical evaporation coef\/f\/icients must have a positive real value when the surface is oriented  along the crystal planes. Since physical restrictions must be independent of orientation of the coordinate axes, it follows from the equivalence transformations that the nonlinear dif\/fusivity $D(y_x)$ must behave like $y_x^{-2}$ at large-slope. This also allows for the existence of a closed non-circular  homothetic solution which cannot exist unless $D(u)$ decreases faster than $u^{-1}$. The Doyle--Vassiliou classif\/ication produces another anisotropic model that satisf\/ies this physical requirement. An exact solution has been constructed, involving Jacobi elliptic functions and other inverse integrals of rational functions.

  Exact solutions sometimes have the advantage of leading to concise conceptually simple relationships. For example, Fig.~\ref{fig3} demonstrates the ef\/f\/icacy of a simple expression for wave amplitude of a corrugated nano-scale surface in the early stages of smoothing by evaporation-condensation when the system cannot be adequately described by a linear model. However, exact solutions can be obtained only in very special cases of initial and boundary conditions, so that approximate numerical solution methods will continue to be important.

\subsection*{Acknowledgements}

This paper is submitted in appreciation of the valuable on-going contributions of  Professor Willard Miller Jr.  The f\/irst author gratefully acknowledges support by the Australian Research Council under project  DP1095044.

\pdfbookmark[1]{References}{ref}
\LastPageEnding

\end{document}